\documentclass[showpacs,amsmath,amssymb]{revtex4}
       
\usepackage{bbold}
\usepackage{epsfig}
\usepackage{rotating}
\usepackage{graphpap}

\begin{document}
\setlength{\unitlength}{1mm}

\title{High-Order Variational Calculation for the Frequency of Time-Periodic Solutions}

\author{Axel Pelster and Hagen Kleinert}

\affiliation{Institute of Theoretical Physics, Free University of Berlin, Arnimallee 14, D--14195 Berlin, Germany\\
E-mail: {\tt pelster@physik.fu-berlin.de, kleinert@physik.fu-berlin.de} }

\author{Michael Schanz}

\affiliation{Institute of Parallel and Distributed Systems (IPVS),
University of Stuttgart, \\ Breitwiesenstra{\ss}e 20-22, D-70565
Stuttgart, Germany\\ E-mail: {\tt
michael.schanz@informatik.uni-stuttgart.de} }

\date{\today}

\begin{abstract}
We develop a convergent variational perturbation theory for the frequency of time-periodic solutions of nonlinear dynamical systems.
The power of the theory is illustrated by applying it to the Duffing oscillator.
\end{abstract}
\maketitle
\section{Introduction}
Perturbative treatments of physical problems provide us with divergent power series in some coupling constant $g$.
Typically, the perturbation coefficients grow factorially, so that they have a zero radius of
convergence. If the signs of the perturbation coefficients alternate, there exist
various resummation schemes which help us to obtain finite results for all values of the
coupling constant $g$, even in the strong-coupling limit $g \rightarrow \infty$ (for an overview see Chap.~16 of 
Ref. \cite{Kleinert2}). Most successful is variational perturbation theory which was recently developed 
\cite{Kleinert1,Kleinert3} as a systematic extension of the variational approach of Feynman and Kleinert \cite{Feynman}. Initially,
this theory was set up for calculating the effective classical potential in quantum statistics.
It has been thoroughly tested for the ground-state energy of the anharmonic oscillator and shown to converge exponentially fast and
uniformly to the correct result \cite{Janke1,Janke2}. This success has led to applications
to divergent series in other branches of theoretical physics \cite{Kleinert1}. Most spectacular was the success in
calculating the most accurate critical exponents of the $\phi^4$-theory without using re-normalization group methods
\cite{Kleinert2,Helium}.\\

In this paper we extend variational perturbation theory by developing 
an exponentially fast converging variational perturbation theory for 
the frequency of time-periodic solutions of nonlinear
dynamical systems. As a simple but nontrivial model  
we consider the one-dimensional anharmonic oscillator with the equation of motion
\begin{eqnarray}
\label{D1}
\ddot{x} ( t ) + \omega_0^2 \,x ( t ) + g \,x^3 ( t )= 0 \, ,
\end{eqnarray}
which is also known as the Duffing equation \cite{Bender}.
Here the dot abbreviates the derivative with respect to the time $t$,
$\omega_0$ denotes the harmonic frequency, and $g>0$ stands for the coupling constant. 
In the following we solve (\ref{D1}) for the initial values
\begin{eqnarray}
\label{I1}
x ( 0 ) = 1 \, , \quad \dot{x} ( 0 ) = 0 \, ,
\end{eqnarray}
and determine the frequency $\omega$ of the resulting periodic motion by using variational perturbation theory.
In Section \ref{sec2} we calculate the frequency $\omega$ as a power series of the coupling constant $g$. Section \ref{sec3}
then elaborates the variational resummation of this weak-coupling series so that the frequency $\omega$ can be determined
for all values of the coupling constant $g$ including the strong-coupling limit $g \rightarrow \infty$.
\section{Perturbation Theory}\label{sec2}
We start by solving the initial value problem (\ref{D1}) and (\ref{I1}) perturbatively to high orders.
\subsection{Poincar\'e-Lindstedt Method}
We assume for a sufficiently small coupling constant $g$ that the solution $x(t)$ has the asymptotic representation
\begin{eqnarray}
\label{EX1}
x ( t ) = x_0 ( t ) + x_1 ( t ) g + \ldots \, .
\end{eqnarray}
A systematic standard procedure to obtain such an asymptotic series for a periodic solution
\begin{eqnarray}
\label{PER1}
x ( t ) = x \left( t+ \frac{2 \pi}{\omega}\right)
\end{eqnarray}
is provided by the Poincar\'e-Lindstedt method \cite{mickens,minorsky}. There one explicitly takes into account that 
the unperturbed frequency $\omega_0$ is shifted to the frequency $\omega$ by a nonzero
coupling constant $g$. One performs a rescaling of time according to 
\begin{eqnarray}
\xi = \omega \, t 
\end{eqnarray}
and introduces the new variable
\begin{eqnarray}
q ( \xi ) = x \left( \frac{\xi}{\omega} \right) \, .
\end{eqnarray}
This converts the periodicity condition (\ref{PER1}) to
\begin{eqnarray}
q ( \xi ) = q \left( \xi + 2 \pi \right) 
\end{eqnarray}
and transforms the original initial value problem (\ref{D1}) and (\ref{I1}) to
\begin{eqnarray}
\label{D2}
\omega^2 \, q'' ( \xi ) + \omega_0^2 \,q ( \xi ) + g \,q^3 ( \xi )= 0 \, , \quad q ( 0 ) = 1 \,, \quad q' (0)=0 \, ,
\end{eqnarray}
where the prime indicates the derivative with respect to the dimensionless new time variable $\xi$.
Since the coupling constant $g$ is supposed to be small, we can expand the frequency
$\omega$ and the period solution $q(\xi)$ in powers of $g$ according to
\begin{eqnarray}
\omega & = & \sum_{n=0}^{\infty} w_n \,\omega_0 \left( \frac{g}{\omega_0^2}\right)^n \, , \label{A1} \\
q ( \xi ) & = & \sum_{n=0}^{\infty} q_n ( \xi )  \left( \frac{g}{\omega_0^2}\right)^n \, .\label{A2}
\end{eqnarray}
Due to this ansatz the expansion coefficients $w_n$ and $q_n(\xi)$ are dimensionless. 
Inserting (\ref{A1}) and (\ref{A2}) in the initial value problem (\ref{D2}) 
and comparing equal powers in the coupling constant $g$ leads for
$n=1,2,\ldots$ to the following recursive set of ordinary differential equations:
\begin{eqnarray}
\label{HY}
q_n'' ( \xi ) + q_n ( \xi )&=& f_n (\xi) \, ,\quad q_n ( 0 ) = q_n' ( 0 ) = 0 \, ,
\end{eqnarray}
where the inhomogeneity $f_n (\xi)$ is given by
\begin{eqnarray}
\label{INHOM}
f_n (\xi) = -2 w_n \,q_0'' ( \xi ) -2 \sum_{l=1}^{n-1} w_l \, q_{n-l}'' (\xi ) 
- \sum_{m=1}^{n-1} \sum_{l=1}^{n-m} w_m \,w_l \,q_{n-m-l}'' (\xi) 
-  \sum_{m=0}^{n-1} \sum_{l=0}^{n-m-1} q_m (\xi) \,q_l (\xi) \,q_{n-m-l-1} ( \xi ) \, .
\end{eqnarray}
This is solved starting from
\begin{eqnarray}
w_0 = 1 \, , \quad q_0 ( \xi ) = \cos \xi \, .
\end{eqnarray}
In the $n$th integration process we proceed in three steps. At first,
we calculate the inhomogeneity $f_n (\xi)$ according to (\ref{INHOM}) 
and expand it in a Fourier series which turns out to be of the following form:
\begin{eqnarray}
f_n (\xi) = \sum_{k=0}^n f_{n,k} \, \cos ( 2 k + 1) \xi \, .
\end{eqnarray}
Second, we prevent a secular term in $q_n (\xi)$ from solving (\ref{HY}) by demanding the condition
\begin{eqnarray}
f_{n,1} = 0 \, ,
\end{eqnarray}
from which the expansion coefficient $w_n$ is uniquely determined. Third, the initial value problem (\ref{HY}) is solved by
\begin{eqnarray}
q_n ( \xi ) = \left[ \sum_{k=1}^n \frac{f_{n,k}}{(2k+1)^2-1} \right] \, \cos \xi + \sum_{k=1}^n \frac{f_{n,k}}{1-(2k+1)^2}
\, \cos (2k+1) \xi \, .
\end{eqnarray}
Using a computer algebra program we obtain in this way the perturbation expansions for both the frequency
\begin{eqnarray}
\label{W0}
\omega = \omega_0  + \frac{3}{8 \omega_0} g - \frac{21}{256 \omega_0^3} g^2 + \ldots
\end{eqnarray}
and the periodic solution
\begin{eqnarray}
x ( t ) &=& \cos \omega t  + \left( - \frac{1}{32 \omega_0^3} \cos \omega t +  \frac{1}{32 \omega_0^3} \cos 3 \omega t
\right) g  \nonumber \\
&& + \left( \frac{23}{1024 \omega^4_0} \cos \omega t  - \frac{3}{128 \omega_0^4}\cos 3 \omega t
+ \frac{1}{1024 \omega^4_0} \cos 5 \omega t \right) g^2 + \ldots \, .
\end{eqnarray}
Tab. \ref{tab1} shows the first 20 weak-coupling coefficients $w_n$ of the frequency $\omega$.

\begin{table}[t]
\begin{center}
\begin{tabular}{|c|c||c|c|} \hline
$n$ & $w_n$        & $n$ & $w_n$ \\[1mm] \hline \hline
\rule[-5pt]{0pt}{23pt}
1 & ${\displaystyle \frac{3}{8}}$     & \hspace*{1mm} 11 \hspace*{1mm}  & 
${\displaystyle \frac{3511276321347}{562949953421312}}$  \\[3mm] \hline
\rule[-5pt]{0pt}{23pt}
2 & ${\displaystyle - \frac{21}{256}}$      & 12  & $-{\displaystyle \frac{401225915283063}{72057594037927936}}$ \\[3mm] \hline
\rule[-5pt]{0pt}{23pt}
3 & ${\displaystyle \frac{81}{2048}}$                 & 13  & 
${\displaystyle \frac{2892201453147555}{576460752303423488}}$ \\[3mm] \hline
\rule[-5pt]{0pt}{23pt}
4 & ${\displaystyle - \frac{6549}{262144}}$    & 14  & 
$-{\displaystyle \frac{84053106665670789}{18446744073709551616}}$ \\[3mm] \hline
\rule[-5pt]{0pt}{23pt}
5 & ${\displaystyle \frac{37737}{2097152}}$  & 15  & 
${\displaystyle \frac{614845335384090729}{147573952589676412928}}$ \\[3mm] \hline
\rule[-5pt]{0pt}{23pt}
6 & \hspace*{1mm} ${\displaystyle -\frac{936183}{67108864}}$  \hspace*{1mm}& 16  & 
$-{\displaystyle \frac{1158192705499996341141}{302231454903657293676544}}$ \\[3mm]\hline
\rule[-5pt]{0pt}{23pt}
7 & ${\displaystyle \frac{6077907}{536870912}}$ & 17 & 
${\displaystyle \frac{8566538482894401288225}{2417851639229258349412352}}$
\\[3mm]\hline
\rule[-5pt]{0pt}{23pt}
8 & $-{\displaystyle \frac{2604833685}{274877906944}}$ & 18 & 
$-{\displaystyle \frac{254612814518190043882263}{77371252455336267181195264}}$
\\[3mm] \hline
\rule[-5pt]{0pt}{23pt}
9 & ${\displaystyle \frac{17839453041}{2199023255552}}$ & 19 & 
${\displaystyle \frac{1899627691040292362960331}{618970019642690137449562112}}$
\\[3mm] \hline
\rule[-5pt]{0pt}{23pt}
10 & \hspace*{1mm} $-{\displaystyle \frac{497158650207}{70368744177664}}$\hspace*{1mm} & 20  & \hspace*{1mm}
$-{\displaystyle \frac{227596989316436230247319519}{79228162514264337593543950336}}$  \hspace*{1mm}
\\[3mm] \hline
\end{tabular}
\end{center}
\caption{\label{tab1}
The first 20 dimensionless weak-coupling coefficients $w_n$ for the frequency $\omega$ of the Duffing oscillator.}
\end{table}
\subsection{Analytical Expression for the Frequency}
\begin{figure}[t]
\centerline{
\setlength{\unitlength}{1mm}
\begin{picture}(100,100)
%
\put(17,  95)   {\footnotesize $\bar 1$}
\put(41.5,95)   {\footnotesize $\bar 3$}
\put(49,  95)   {\footnotesize $\bar 5$}
\put(52.5,95)   {\footnotesize $\bar 7$}
\put(55,  95)   {\footnotesize $\bar 9$}
\put(61.2,95)   {\footnotesize $\underline 9$}
\put(63,  95)   {\footnotesize $\underline 7$}
\put(65.5,95)   {\footnotesize $\underline 5$}
\put(68,  95)   {\footnotesize $\underline 3$}
\put(70,  95)   {\footnotesize $\underline 1$}
\put(62.5, 9)     {\footnotesize $\underline 8$}
\put(65,   9)     {\footnotesize $\underline 6$}
\put(69.5, 9)     {\footnotesize $\underline 4$}
\put(79.5, 9)     {\footnotesize $\underline 2$}
\put(46.5, 9)     {\footnotesize $\bar 2$}
\put(50.5, 9)     {\footnotesize $\bar 4$}
\put(53.2, 9)     {\footnotesize $\bar 6$}
\put(55.3, 9)     {\footnotesize $\bar 8$}
\put(48,0){$\lg~g$}
\put(0,48){\begin{sideways}$\omega(g)$\end{sideways}}
\put(0,5){\epsfig{figure=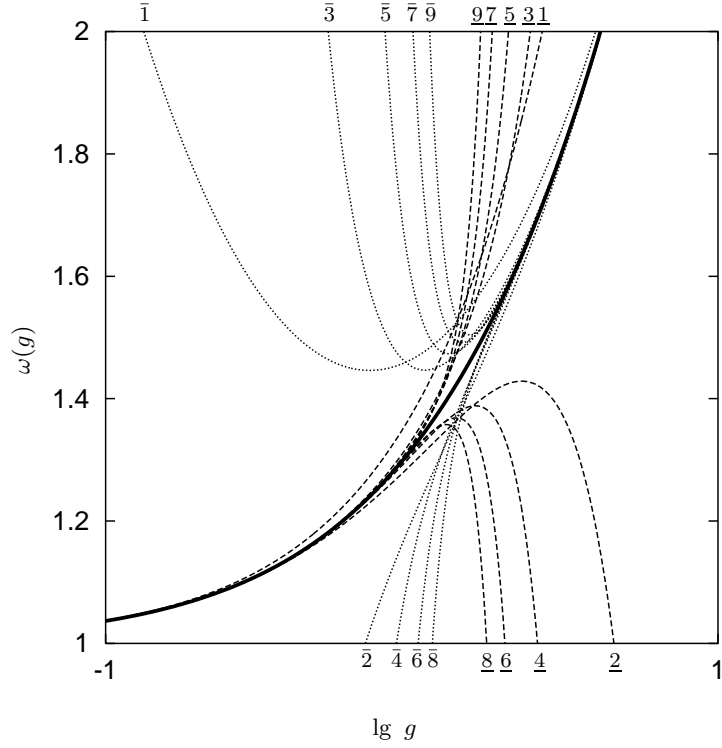, width=100\unitlength}}
\end{picture}}
\caption{\label{weak/strong} Logarithmic plot of the full function
(\ref{ANA}) (solid curve) versus the coupling constant $g$ compared
with the truncated successive divergent perturbation expansions
(\ref{A1}) (dashed curves, the corresponding orders are labeled with
$\underline{1} \ldots \underline{9}$) and the partial sums of the
convergent strong-coupling expansions (\ref{SC1}) (dotted curves, the
corresponding orders are labeled with $\bar{1} \ldots \bar{9}$).}
\end{figure}

Remarkably,
the frequency $\omega$ of the Duffing oscillator (\ref{D1}) can be determined exactly as a function of the coupling constant
$g$. To this end we multiply (\ref{D1})
by $\dot{x}(t)$ and integrate once, taking into account the initial values (\ref{I1}): 
\begin{eqnarray}
\label{EC}
\frac{1}{2} \dot{x}^2 ( t ) + \frac{1}{2} \omega_0^2 \, x^2 ( t ) + \frac{1}{4} g \, x^4 ( t ) =
\frac{1}{2} \, \omega_0^2   + \frac{1}{4} \, g  \, ,
\end{eqnarray}
Separating the variables in the energy conservation (\ref{EC}) and integrating over a quarter of the 
oscillator period, we get
\begin{eqnarray}
\frac{\pi}{2 \,\omega} = \int\limits_0^{1} \frac{d x}{\sqrt{\omega_0^2 + {\displaystyle \frac{g}{2}}  - 
\omega_0^2 x^2 - {\displaystyle \frac{g}{2}} x^4 } } \, .
\end{eqnarray}
The integral can be explicitly performed by using Eq. (3.152.4) in Ref. \cite{gradshteyn}
\begin{eqnarray}
\label{ANA}
\omega = \frac{\pi \sqrt{\omega_0^2 + g} }{2 \,F \left( {\displaystyle \frac{\pi}{2}} , 
\sqrt{{\displaystyle \frac{g}{2 (\omega_0^2 + g)}} } \,\, \right) } \, ,
\end{eqnarray}
where $F$ denotes the elliptic integral of the first kind which is defined in Eq. (8.111.2) of Ref. \cite{gradshteyn}:
\begin{eqnarray}
F(\varphi,k) = \int\limits_0^{\varphi} \, \frac{d \alpha}{\sqrt{1 - k^2 \sin^2 \alpha}} \, .
\end{eqnarray}
In the weak-coupling limit $g \rightarrow 0$, we recover from (\ref{ANA}) the perturbation series (\ref{A1})
of the previous section by taking into account (8.113.1) of  Ref. \cite{gradshteyn}.
However, the exact result (\ref{ANA}) also allows to find the strong-coupling limit $g \rightarrow \infty$:
\begin{eqnarray}
\label{SC1}
\omega  = \sqrt{g} \, \left[ b_0 + b_1 \frac{\omega_0^2}{g} + 
b_2 \left( \frac{\omega_0^2}{g} \right)^2 + \ldots \right] \quad (g \rightarrow \infty) \, .
\end{eqnarray}
The leading strong-coupling coefficient $b_0$ has the numerical value
\begin{eqnarray}
\label{SC3}
b_0 = \frac{\pi}{\displaystyle 2\, F \left( \frac{\pi}{2}, \frac{1}{\sqrt{2}} \right)} =  
0.8472130847939790866\ldots   \, .
\end{eqnarray}
In Figure \ref{weak/strong} we compare the full function (\ref{ANA}) with the successive
divergent weak-coupling expansions (\ref{A1}) and with the convergent strong-coupling expansions (\ref{SC1}).
We observe that the full function represents the envelope to all weak- and strong-coupling expansions.
\section{Variational Perturbation Theory}\label{sec3}
Let us now see how well we can reproduce the strong-coupling result by
resumming the weak-coupling series (\ref{A1}) of the frequency $\omega$ with the help of variational perturbation
theory \cite{Kleinert1,Kleinert2}. In this way we are able obtain approximations of the frequency $\omega$ 
in good agreement with the exact result for all values of the coupling constant $g$. In particular, we carry 
the strong-coupling limit $g \rightarrow \infty$ of the frequency $\omega$ to high orders in order to investigate in detail
the convergence of the variational results. 
\subsection{General Procedure}
We start with the weak-coupling expansion (\ref{A1}) truncated at order $N$:
\begin{eqnarray}
\label{W1} 
\omega^{(N)} =  \sum_{n=0}^N w_n \,\omega_0^{1-2n} g^n \, . 
\end{eqnarray}
The we introduce the variational parameter $\Omega$ by Kleinert's square root trick \cite{Kleinert1}
\begin{eqnarray}
\label{SR}
\omega_0 = \sqrt{\Omega^2 + \omega_0^2 - \Omega^2} = \Omega \sqrt{1+g\,r} \, ,
\end{eqnarray}
where the abbreviation $r$ is defined by
\begin{eqnarray}
\label{ABB}
r = \frac{1}{g} \left( \frac{\omega_0^2}{\Omega^2} -1 \right) \, ,
\end{eqnarray}
and we ignore for the moment that $r$ depends on $g$ and regard it as a constant.
Substituting (\ref{SR}) into the truncated weak-coupling series (\ref{W1}), we obtain
\begin{eqnarray}
\label{W2}
\omega^{(N)} ( g , \Omega) = \sum_{n=0}^N w_n \Omega^{1-2n} \left( 1 + g \, r \right)^{1/2-n} \, g^n \, .
\end{eqnarray}
The factor $(1+g\,r)^{\alpha}$ with $\alpha=1/2-n$ is then expanded up to the order $N-n$
\begin{eqnarray}
\label{BE}
(1+g\,r)^{\alpha}= \sum_{k=0}^{N-n} \left( \begin{array}{@{}c} \alpha \\ k \end{array} \right)
\, \left( g \, r \right)^k + {\cal O} \left( g^{N-n+1}\right) \, ,
\end{eqnarray}
where the binomial coefficient is defined by the Gamma function $\Gamma$:
\begin{eqnarray}
\left( \begin{array}{@{}c} \alpha \\ k \end{array} \right) = \frac{\Gamma(\alpha + 1)}{\Gamma(k+1)\,\Gamma(\alpha+k+1)} \, .
\end{eqnarray}
Thus the sum (\ref{W2}) is re-expanded including all powers of $g$ up to the order $g^N$: 
\begin{eqnarray}
\label{W3}
\omega^{(N)} ( g , \Omega) = \sum_{n=0}^N w_n \Omega^{1-2n} \left[ \,\sum_{k=0}^{N-n} 
\left( \begin{array}{@{}c} 1/2 - n \\ k \end{array} \right) \, \left( \frac{\omega_0^2}{\Omega^2} - 1 \right)^k \,\right] \, g^n \, .
\end{eqnarray}
At the end we have inserted (\ref{ABB}). \\

If we could consider the limit $N \rightarrow \infty$ in (\ref{W3}), the dependence on the
artificially introduced variational parameter $\Omega$ would drop out. Due to the 
truncation at the finite order $N$, however, we obtain an explicit dependence on the variational parameter $\Omega$
in (\ref{W3}). This suggests to fix the yet undetermined variational parameter $\Omega$ according to the principle of 
minimal sensitivity \cite{stevenson}. Thus we try to find at first an extremum:
\begin{eqnarray}
\label{C1}
\left. \frac{\partial \omega^{(N)} (g,\Omega)}{\partial \Omega} \right|_{\Omega=\Omega^{(N)}(g)} = 0 \, .
\end{eqnarray}
If this equation has no real solution $\Omega^{(N)}(g)$, then we look for a saddle point instead \cite{Kleinert1}
\begin{eqnarray}
\label{C2}
\left. \frac{\partial^2 \omega^{(N)} (g,\Omega)}{\partial \Omega^2} \right|_{\Omega=\Omega^{(N)}(g)} = 0 \, ,
\end{eqnarray}
or more generally, for a real zero $\Omega^{(N)}(g)$ of the lowest derivative with respect to the variational parameter.
This optimal value $\Omega^{(N)}(g)$ then leads via
\begin{eqnarray}
\omega^{(N)} ( g ) = \omega^{(N)} \left( g , \Omega^{(N)}(g) \right) 
\end{eqnarray}
to an approximation of the frequency $\omega$ which turns out to lead to good results for all values of the coupling constant $g$.
To lowest order, the optimal solution is unique. If there are several optimal solutions, we always choose the one which
is closest to the optimal solution of the previous order.
\subsection{Lowest Orders}
\begin{figure}[t]
\centerline{
\setlength{\unitlength}{0.8mm}
\begin{picture}(105,75)
\put(0,20)
{\begin{sideways}$\omega(g), \omega^{(1)}(g), \omega^{(2)}(g)$\end{sideways}}
\put(56,0){$g$}
\put(5, 75){\epsfig{ figure=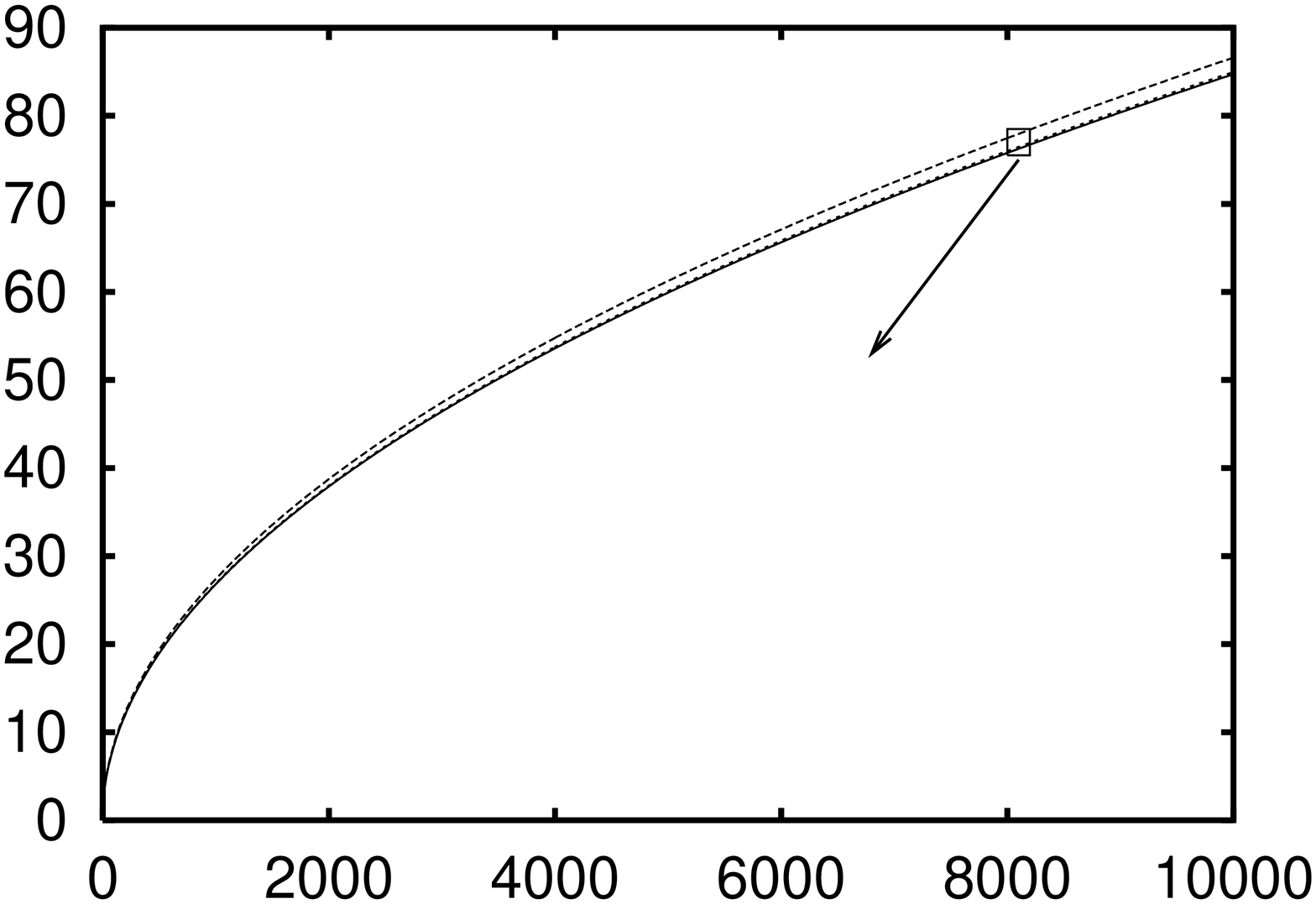, angle=270, width=100\unitlength}}
\put(48, 48){\epsfig{ figure=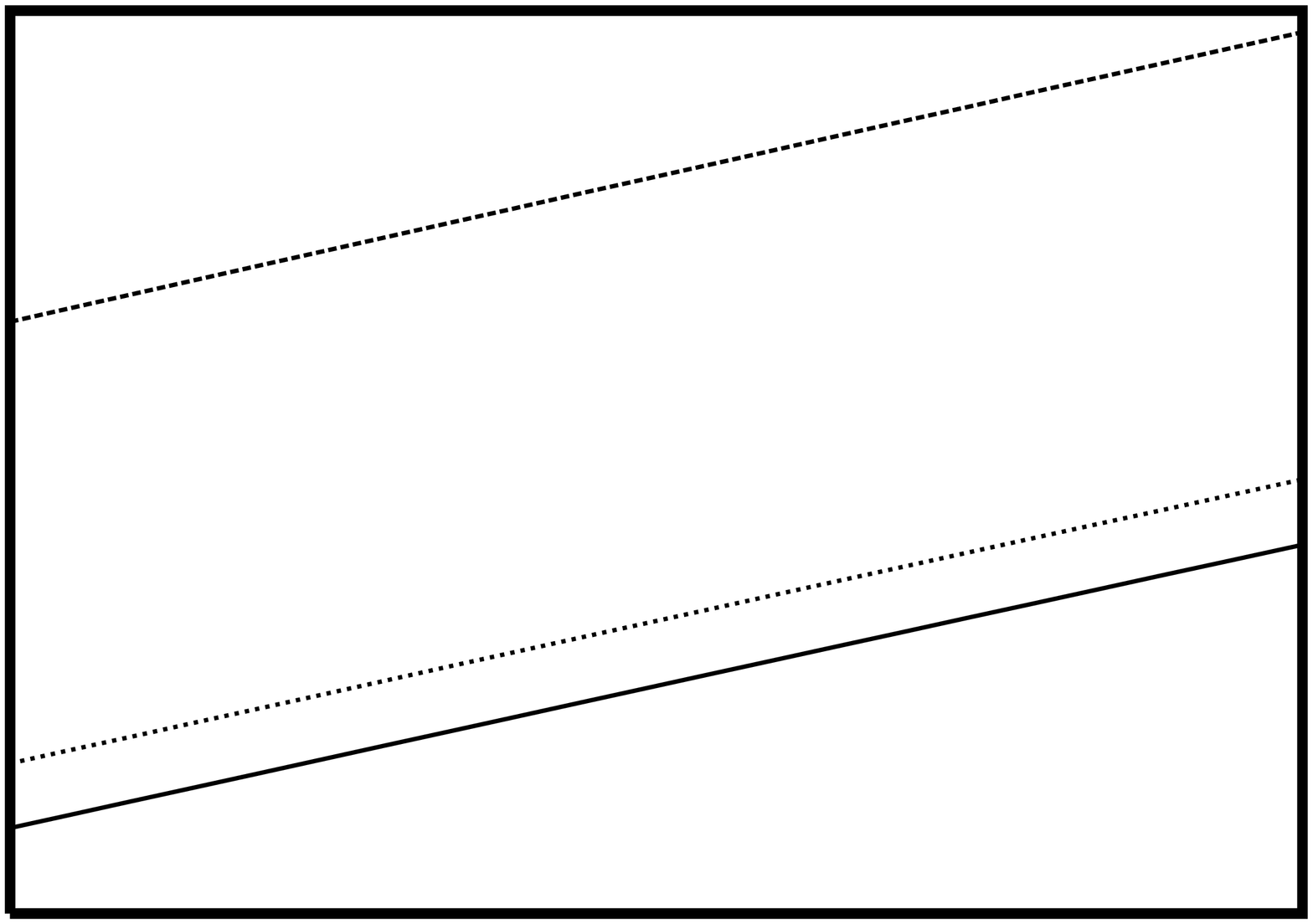, angle=270, width=50\unitlength}}
\put(52.5,40.5){\tiny $\omega^{(1)}(g)$}
\put(52.5,26){\tiny $\omega^{(2)}(g)$}
\put(82,21.5){\tiny $\omega(g)$}
\end{picture}}
\caption{\label{numerics} The first- and second-order variational approximation (\ref{B4}) and (\ref{E4})
for the frequency $\omega$ is compared with the exact result (\ref{ANA}).}
\end{figure}

Let us consider the lowest order approximations of variational perturbation theory explicitly. Truncating the weak-coupling
series (\ref{W0}) at the order $N=1$ leads to
\begin{eqnarray}
\label{B1}
\omega^{(1)} = \omega_0 + \frac{3}{8 \omega_0} \, g \, .
\end{eqnarray}
Applying the general procedure as described in detail in the previous subsection, we obtain
\begin{eqnarray}
\label{B2}
\omega^{(1)} ( g , \Omega) = \frac{\Omega}{2} + \left( \frac{\omega_0^2}{2} + \frac{3 g}{8} \right) \, \frac{1}{\Omega} \, .
\end{eqnarray}
From the condition (\ref{C1}) we determine the variational parameter as
\begin{eqnarray}
\label{B3}
\Omega^{(1)} (g) = \omega_0 \, \sqrt{1 + \frac{3 g}{4 \omega_0^2}} \, ,
\end{eqnarray}
whose substitution into (\ref{B2}) leads to the first-order approximation
\begin{eqnarray}
\label{B4}
\omega^{(1)} ( g ) = \Omega^{(1)} (g) = \omega_0 \, \sqrt{1 + \frac{3 g}{4 \omega_0^2}} \, .
\end{eqnarray}
In a similar way we proceed for the second order, where the truncated weak-coupling series (\ref{W0}) reads
\begin{eqnarray}
\label{E1}
\omega^{(2)} = \omega_0  + \frac{3}{8 \omega_0} \, g - \frac{21}{256 \omega_0^3} \, g^2
\end{eqnarray}
Applying the square root trick leads here to
\begin{eqnarray}
\label{E2}
\omega^{(2)} ( g , \Omega ) = \frac{3}{8} \Omega + \left( \frac{3 \omega_0^2}{4} + \frac{9 g}{16} \right) \frac{1}{\Omega}
- \left( \frac{\omega_0^4}{8} + \frac{3 \omega_0^2 g}{16} + \frac{21 g^2}{256} \right) \frac{1}{\Omega^3} \, .
\end{eqnarray}
It turns out that $\omega^{(2)} ( g , \Omega )$ has no real extremum with respect to $\Omega$. Thus we have to look for a
turning point instead. From condition (\ref{C2}) we obtain
\begin{eqnarray}
\label{E3}
\Omega^{(2)} ( g ) = \omega_0 \, 
\sqrt{\frac{\displaystyle 1 + \frac{3 g}{2 \omega_0^2} + \frac{21 g^2}{32 \omega_0^4} }{\displaystyle 1 + \frac{3 g}{4 \omega_0^2} }}
\, .
\end{eqnarray}
Inserting (\ref{E3}) in (\ref{E2}) leads to the second-order approximation
\begin{eqnarray}
\label{E4}
\omega^{(2)} ( g )= \omega_0 \, \frac{\displaystyle 1 + \frac{3 g}{\omega_0^2} + \frac{153 g^2}{256 \omega_0^4} 
}{\sqrt{\displaystyle \left( 1 + \frac{3 g}{4 \omega_0^2} \right) \left( 1 + \frac{3 g}{2 \omega_0^2} + 
\frac{21 g^2}{32 \omega_0^4} \right)}} \, .
\end{eqnarray}
In Fig.~\ref{numerics} we compare the first- and second-order variational approximation (\ref{B4}) and (\ref{E4})
for the frequency $\omega$ with the exact result (\ref{ANA}). Notably, the first-order variational result is very good
for all values of the coupling constant $g$, and the second-order leads to a substantial improvement of the accuracy.
\subsection{Strong-Coupling Limit}
In order to quantify the accuracy of the variational approximations (\ref{B4}) and (\ref{E4}), 
we study now in particular the strong-coupling regime
$g \rightarrow \infty$. In first and second order we reproduce the expansion (\ref{SC1}) with the leading strong-coupling
coefficient
\begin{eqnarray}
\label{SC2}
b_0^{(1)} = \frac{\sqrt{3}}{2} \approx 0.866025 \, , \quad b_0^{(2)} = \frac{51 \sqrt{14}}{224} \approx 0.851895 \,.
\end{eqnarray}
Comparing (\ref{SC3}) with (\ref{SC2}) we see that the first and the second order of variational perturbation theory yields the
strong-coupling coefficient (\ref{SC3}) within the accuracy of 2.2 \% and 0.55 \%, respectively.\\

\begin{figure}[t]
\centerline{
\setlength{\unitlength}{0.8mm}
\begin{picture}(100,100)
\put(0,40){\begin{sideways}$\ln \left( |b^{(N)}_0 - b_0|/b_0 \right)$\end{sideways}}
\put(53,0){$N$}
\put(0,5){\epsfig{figure=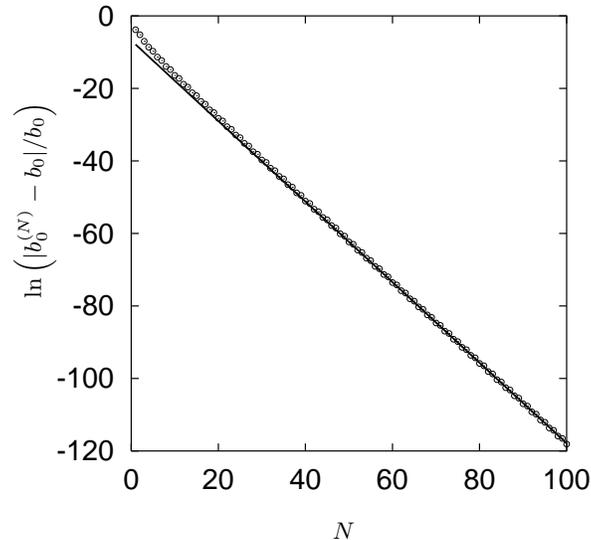, width=100\unitlength}}
\end{picture}}
\caption{\label{log} The points show the logarithmic plot of the error $|b^{(N)}_0 - b_0|/b_0$ against the order 
$N$, and the solid line represents a fit of the last 10 points to the straight line $- \alpha - \beta N$.}
\end{figure}

In order to obtain higher-order variational results for this strong-coupling coefficient (\ref{SC3}), we proceed as follows.
From the first- and second-order approximations (\ref{B3}) and (\ref{E3}) we see that the variational parameter has a
strong-coupling expansion of the form
\begin{eqnarray}
\label{SC4}
\Omega^{(N)} ( g ) = \sqrt{g} \left( \Omega_0^{(N)} + \frac{\Omega_1^{(N)}}{g}+ \frac{\Omega_2^{(N)}}{g^2} + \ldots \right) \, .
\end{eqnarray}
Inserting (\ref{SC4}) in (\ref{W3}), we obtain the $N$th order approximation for (\ref{SC3}) with
\begin{eqnarray}
b_0^{(N)} (\Omega_0^{(N)}) = \sum_{n=0}^N \sum_{k=0}^{N-n} \left( \begin{array}{@{}c} 1/2-n \\ k \end{array} \right)
\, (-1)^k \, w_n \,\left( \Omega_0^{(N)} \right)^{1-2n} \, .
\end{eqnarray}
The inner sum can be done by using Eq. (0.151.4) in Ref. \cite{gradshteyn}:
\begin{eqnarray}
b_0^{(N)} (\Omega_0^{(N)}) =\sum_{n=0}^N  (-1)^{N-n} \, \left( \begin{array}{@{}c} -1/2-n \\ N-n \end{array} \right)
\, w_n \, \left(\Omega_0^{(N)} \right)^{1-2n}\, .
\end{eqnarray}
In order to optimize the variational parameter $\Omega_0^{(N)}$ we look again for an extremum
\begin{eqnarray}
\frac{\partial b_0^{(N)} (\Omega_0^{(N)})}{\partial \Omega_0^{(N)}} = 0 
\end{eqnarray}
or for a saddle point
\begin{eqnarray}
\frac{\partial^2 b_0^{(N)} (\Omega_0^{(N)})}{\partial \Omega_0^{(N)} {}^2} = 0 \, .
\end{eqnarray}
It turns out that an extremum exists for odd orders $N$, whereas even orders $N$ lead to a saddle point.
Tab. \ref{strong} shows the first 20 variational results for the strong-coupling coefficient (\ref{SC3}).
Note that the approximant $b_0^{(20)}$ coincides already in 11 digits with (\ref{SC3}). The points of Figure \ref{log}
show that the logarithm of the error $|b^{(N)}_0 - b_0|/b_0$ depends linearly on $N$
up to the order $N=100$ according to
\begin{eqnarray}
\label{EXP}
\frac{|b^{(N)}_0 - b_0|}{b_0} = e^{- \alpha - \beta \, N }\, ,
\end{eqnarray}
where the fit of the last 10 points leads to the quantities $\alpha =
-6.7671$ and $\beta =-1.1113$. Thus we have demonstrated that the
variational approximations for the frequency of the Duffing oscillator
converge exponentially fast.  Note that the speed of convergence is
considerably faster than the exponential convergence of the
variational results for the ground-state energy of the anharmonic
oscillator \cite{Janke1,Janke2}.
\begin{table}[t]
\begin{center}
\begin{tabular}{|c|c||c|c|} \hline
$N$ & $b_0^{(N)}$        & $N$ & $b_0^{(N)}$ \\[1mm] \hline \hline
\rule[-5pt]{0pt}{20pt}
1 & \hspace*{1mm} $0.86602540378443864676$ \hspace*{1mm} & \hspace*{1mm}11 \hspace*{1mm}& 
 \hspace*{1mm} $0.84721311260106078088$  \hspace*{1mm}  \\[2mm] \hline
\rule[-5pt]{0pt}{20pt}
2 & \hspace*{1mm} $0.85189520859585272618$   \hspace*{1mm}    & 12  & $0.84721309038427087031$ \\[2mm] \hline
\rule[-5pt]{0pt}{20pt}
3 & $0.84798320787226284162$                 & 13  & $0.84721308733437656102$ \\[2mm] \hline
\rule[-5pt]{0pt}{20pt}
4 & $0.84736735286736694523$    & 14  & $0.84721308530703137833$ \\[2mm] \hline
\rule[-5pt]{0pt}{20pt}
5 & $0.84726277296604748829$  & 15  & $0.84721308503241446175$ \\[2mm] \hline
\rule[-5pt]{0pt}{20pt}
6 & $0.84722291812428697005$  \hspace*{1mm}& 16  & $0.84721308484231654612$ \\[2mm]\hline
\rule[-5pt]{0pt}{20pt}
7 & $0.84721687569394258505$ & 17 & $0.84721308481682089873$
\\[2mm]\hline
\rule[-5pt]{0pt}{20pt}
8 & $0.84721383828896139276$ & 18 & $0.84721308479862454760$
\\[2mm] \hline
\rule[-5pt]{0pt}{20pt}
9 & $$0.84721340071349571092 & 19 & $0.84721308479620273029$
\\[2mm] \hline
\rule[-5pt]{0pt}{20pt}
10 \hspace*{1mm}& $0.84721314796371865932$& 20  & $0.84721308479443254139$  
\\[2mm] \hline
\end{tabular}
\end{center}
\caption[The first 20 weak-coupling coefficients for the instanton problem]
{\label{strong}
The first 20 variational results for the strong-coupling coefficient (\ref{SC3}).}
\end{table}
\section{Conclusion}
We have demonstrated by the example of the Duffing oscillator how variational perturbation theory is
successively applied to determine the frequency of time-periodic solutions of nonlinear dynamical systems.
It remains to proceed along similar lines to treat also nonconservative systems with limit cycles like 
the van der Pol equation \cite{Bender}. 
\section*{Acknowledgment}
One of us,
A.P., is grateful for the hospitality of G\"unter Wunner at the I. Institute of Theoretical Physics
at the University of Stuttgart where this article was finished.


\begin{thebibliography}{199}
%
\bibitem{Kleinert2} 
H.~Kleinert and V.~Schulte-Frohlinde, {\it Critical Properties of $\phi^4$--Theories}
(World Scientific, Singapore, 2001).
%
\bibitem{Kleinert1} 
H.~Kleinert, {\it Path Integrals in Quantum Mechanics,
Statistics, and Polymer Physics}, second edition
(World Scientific, Singapore, 1995).
%
\bibitem{Kleinert3}
H.~Kleinert, {\it Phys.~Lett.~A} {\bf 173}, 332 (1993).
%
\bibitem{Feynman}
R.~P.~Feynman and H.~Kleinert, {\it Phys.~Rev.~A} {\bf 34}, 5080 (1986).
%
\bibitem{Janke1}
W.~Janke and H.~Kleinert, {\it Phys.~Rev.~Lett.} {\bf 75}, 2787 (1995).
%
\bibitem{Janke2}
H.~Kleinert and W.~Janke, {\it Phys.~Lett.~A} {\bf 206}, 283 (1995).
%
\bibitem{Helium}
H. Kleinert, {\it Phys.~Lett.~A} {\bf 277}, 205 (2000).
%
\bibitem{Bender}
C.M. Bender and S.A. Orszag, {\it Advanced Mathematical Methods for Scientists and Engineers --
Asymptotic Methods and Perturbation Theory} (McGraw-Hill, New York, 1978).
%
\bibitem{mickens}
R. Mickens, {\it Introduction to Nonlinear Oscillations} (Cambridge University Press, Cambridge, 1981).
%
\bibitem{minorsky}
N. Minorsky, {\it Nonlinear Oscillation} (Van Nostrand, Princeton, 1962).
%
\bibitem{gradshteyn}
I.S.~Gradshteyn and I.M.~Ryzhik, {\it Table of Integrals, Series, and Products},
corrected and enlarged edition (Academic Press, New York, 1980).
%
\bibitem{stevenson}
P.~M.~Stevenson, {\it Phys.~Rev.~D} {\bf 23}, 2916 (1981).
%
\end{thebibliography}
\end{document}